\documentclass[preprintnumbers,amsmath,amssymb,twocolumn]{revtex4}
\usepackage{graphicx}
\usepackage{dcolumn}
\usepackage{bm}
\usepackage{verbatim}
\usepackage{comment}
\usepackage{epsfig}
\usepackage{color}
\usepackage{multirow}
\usepackage{ulem}

\renewcommand{\[}{\begin{equation}}
\renewcommand{\]}{\end{equation}}

\renewenvironment{thebibliography}[1]
        {\begin{list}{\arabic{enumi}.}
        {\usecounter{enumi}\setlength{\parsep}{0pt}
         \setlength{\itemsep}{0pt}
         \settowidth
        {\labelwidth}{#1.}\sloppy}}{\end{list}}

\begin{document}
\title{Novel technique for supernova detection with IceCube}
\author{L. Demir\"ors, M. Ribordy and M. Salathe} 
\address{High Energy Physics Laboratory, EPFL, CH-1015 Lausanne, Switzerland}

\begin{abstract}


The current supernova detection technique used in IceCube relies
on the sudden deviation of the summed photomultiplier noise rate from
its nominal value during the neutrino burst, making IceCube a $\approx 3$
Megaton effective detection volume - class supernova detector. While
galactic supernovae can be resolved with this technique, the supernova 
neutrino emission spectrum remains unconstrained and thus
presents a limited potential for the topics related to  
supernova core collapse models.

The paper elaborates analytically on the capabilities of IceCube to
detect supernovae through the analysis of hits in the detector correlated 
in space and time. These arise from supernova neutrinos interacting in the
instrumented detector volume along single strings. Although the effective
detection volume for such coincidental hits is much smaller ($\gtrsim
35\,$kton, about the scale of SuperK), a wealth of information is
obtained due to
the comparatively low coincidental noise rate.
We demonstrate that a neutrino flux from a core collapse
supernova will produce a signature enabling the resolution of rough spectral
features and, in the case of a strong signal, providing indication on its
location.

We further discuss the enhanced potential of a rather modest
detector extension, a denser array in the center of IceCube, within
our one dimensional analytic calculation framework. Such an extension
would enable the exploration of the neutrino sky above a few GeV
and the detection of supernovae up to a few 100's of kilo parsec. However, 
a $3-4\,$Mpc detection distance, necessary for routine supernova detection,
demands a significant increase of the effective detection volume and can be 
obtained only with a more ambitious instrument, particularly the boosting of 
sensor parameters such as the quantum efficiency and light collection area. 

\end{abstract}



\maketitle

\section{Introduction}
Much of the underlying physics of a core collapse supernova remains unexplored, the mechanism leading to the explosion not well
understood. Though it is clear that neutrinos play a
major role in the dynamics of the explosion, carrying away 99\% of the
energy emitted by the dying star. The
observation of neutrinos in the provenance of SN 1987A in several experiments
\cite{kamiokandeSN1987A,IMBSN1987A,baksanSN1987A} has confirmed the crucial 
role they play in the overall picture. The limited number of detected neutrinos, 
however, has not provided enough statistics to constrain current
supernova models. This may only be achieved with a detector performing
routine SN detection, about one per year.

The idea of using IceCube as a supernova detector was first discussed
in~\cite{pryor, halzen-sn}. The method is based on the sudden increase
in the photomultiplier count rate on a timescale of the order of $10\,$seconds:
qualitatively, the release by a supernova at $d=10\,$kpc with a total
amount of energy of $10^{58}\,$MeV in neutrinos of about $20\,$MeV
corresponds to a flux of $7\cdot 10^{14}\,\mathrm{m}^{-2}$
In its completed
configuration and with its data acquisition system, IceCube has about $3\,$Mton
effective detection volume and a reach of about $50\,$kpc at
$5\sigma$ C.L.\ detection level. Here, we extend that method and
assess the potential of IceCube and several prospective detector
configurations by considering multiple hit detection in a single or
neighboring sensors from individual neutrino interactions. 
We demonstrate that multi-hit modes enable the extraction of new observables
related to the features of the neutrino emission spectrum and the supernova location,
not accessible by means of the original method.
This new data stream could help study questions related to
supernova dynamics (neutronization burst, accretion and cooling phase,
see \cite{Janka:SN} and references therein) or fundamental
questions of particle physics, related to the nature of the
neutrino \cite{Raffelt:physics,Dighe:physics}. The
observation of electron neutrinos from the
early neutronization burst 
could disentangle the neutrino hierarchy \cite{Dighe:hierarchy}, 
provided a sufficiently large mixing angle $\theta_{13}$ as indicated 
recently \cite{Abe:2011sj}.

These new detection modes are conceptually situated between a
full reconstruction of the events and the current search method balancing
the simplicity of the latter with the richness in information of the
former. We show that the loss of information in the current method is
not necessary and that a very elementary reconstruction, implemented through
the coincident hit methodology, can enhance the information obtained
during a supernova explosion. In fact, it would appear natural to
extend the new detection modes to a full reconstruction of supernova
events in the case of a very densely instrumented volume.

According to~\cite{AndoAndBeacom, KistlerAndBeacom}, the estimated SN rate from
our galaxy is about $0.03/\mathrm{yr}$. It increases to
$0.1/\mathrm{yr}$ within $1\,$Mpc and about $1/\mathrm{yr}$ within
$4\,$Mpc. We will see that extending the reach of a future detector to
such a large distance is quite challenging given the quick decrease of
the interaction density with distance.
SuperK, with its $32\,$kton fiducial volume, remains at the moment an
instrument of choice \cite{superk}. IceCube \cite{IceCube} with its 86
strings and this new detection methodology could become a serious competitor,
even surpass it, provided a deep and denser
core extension installation. There are several good reasons for such an extension
beyond an enhanced potential for SN detection, mainly the detection
of individual high energy neutrinos down to several GeV related to
multiple physics aspects, such as neutrino sources in the Southern sky
\cite{Roucelle:SouthernSky,Sestayo:SouthernSky}, indirect detection of
dark matter down to lower mass
\cite{Abbasi:2009vg,Heros:2010ss,Abbasi:2011eq} and neutrino particle
physics \cite{DeepCore:physics}.

This contrasts with the potential of other projected $100\,$kton-class detectors, e.g.\ liquid scintillator LENA~\cite{LENA} 
or liquid Argon detectors~\cite{rubbia}, which will have an extended potential for SN detection but comparatively small fiducial volumes at energies beyond GeV. 
Especially the latter are complementary to water or ice-based Cherenkov detectors, 
since they are more sensitive to the electron neutrino channel, providing a
better access to the physics during the neutronization burst
\cite{Choubey}. Combining the measurements from these detectors, which
differ not only in their sensitive material, but in their
geographic location, would enable the study of matter effects and
neutrino hierarchy by observing neutrinos from the same supernova from
different viewpoints with a high detection probability.

Finally, a large detector will not only increase the distance reach
for SN but also provide large event statistics for galactic SN with
possible resolution for their location, leading through a SN
alert system (SNEWS~\cite{SNEWS}) to the optical detection of
obscured and otherwise missed galactic SNe.

In Section 2, we define the studied
configuration and the benchmark supernova neutrino flux. 
We present the effective volume analytical
calculations in Section 3 and their consequences in Section 4.

\section{definitions and methodology}
\subsection{Detectors and sensors}

\vskip3mm
{\noindent\bf Configurations}
\vskip1mm
\noindent 
In the following, we consider several detector
configurations~\cite{Koskinen:IceCube}:
\paragraph{The baseline IceCube configuration} (IC), with a number  of
sensors $N_{\rm{OM}}=4800$, arranged $L=17\,$m apart along 80
strings. 
\paragraph{The Deep Core configuration} (DC), with $N_{\rm{OM}}=320$,
$L=7\,$m apart on 8 strings. 
\paragraph{The Dense and Deep Core configuration:} An imaginative
futuristic low energy extension in  alternative configurations denoted
DDC, DDC$_{4\pi}$, DDC$_{4\pi}^{\rm{VL}}$ composed of 24 strings with
150 sensors each, $N_{\rm{OM}}=3600$, and with a spacing between
modules of $L=3\,$m. In configurations with ${4\pi}$ subscript, the
sensitive area of the spherical module has a $4\pi$ acceptance, for
instance the modules designed for KM3NET~\cite{km3net-4piDOM}. The
DDC$_{4\pi}^{\rm{VL}}$ configuration has also a larger sensitive area
and higher Q.E. In addition, we assess the potential of the full
acceptance configurations with extremely low noise modules ($100\,$Hz).

A short local coincidence is the occurrence of two or more hits within
specific time windows $\delta t$'s in a single sensor or in two
neighboring sensors, also referred to as optical modules. The
systematic recording of these coincidences would require the
implementation of a dedicated data acquisition mode and its
corresponding data stream. We study in this paper several detection
modes $ij$, $i\le2$ and $j\le1$, referring to the coincidental
detection of $i$ hits with a module and $j$ hits with the module just
above along a sensor string. 

In this paper we do not consider coincident hits between strings,
so string spacing is not a parameter in this study. We refer
therefore to our study as 1D analytic approximation. We
also restrict the study to nearest neighbor (NN) coincidences. 

\begin{figure}[h]
\centering
 \includegraphics*[width=0.4\textwidth,angle=0,clip,trim=0 16.5cm 7.5cm 0]{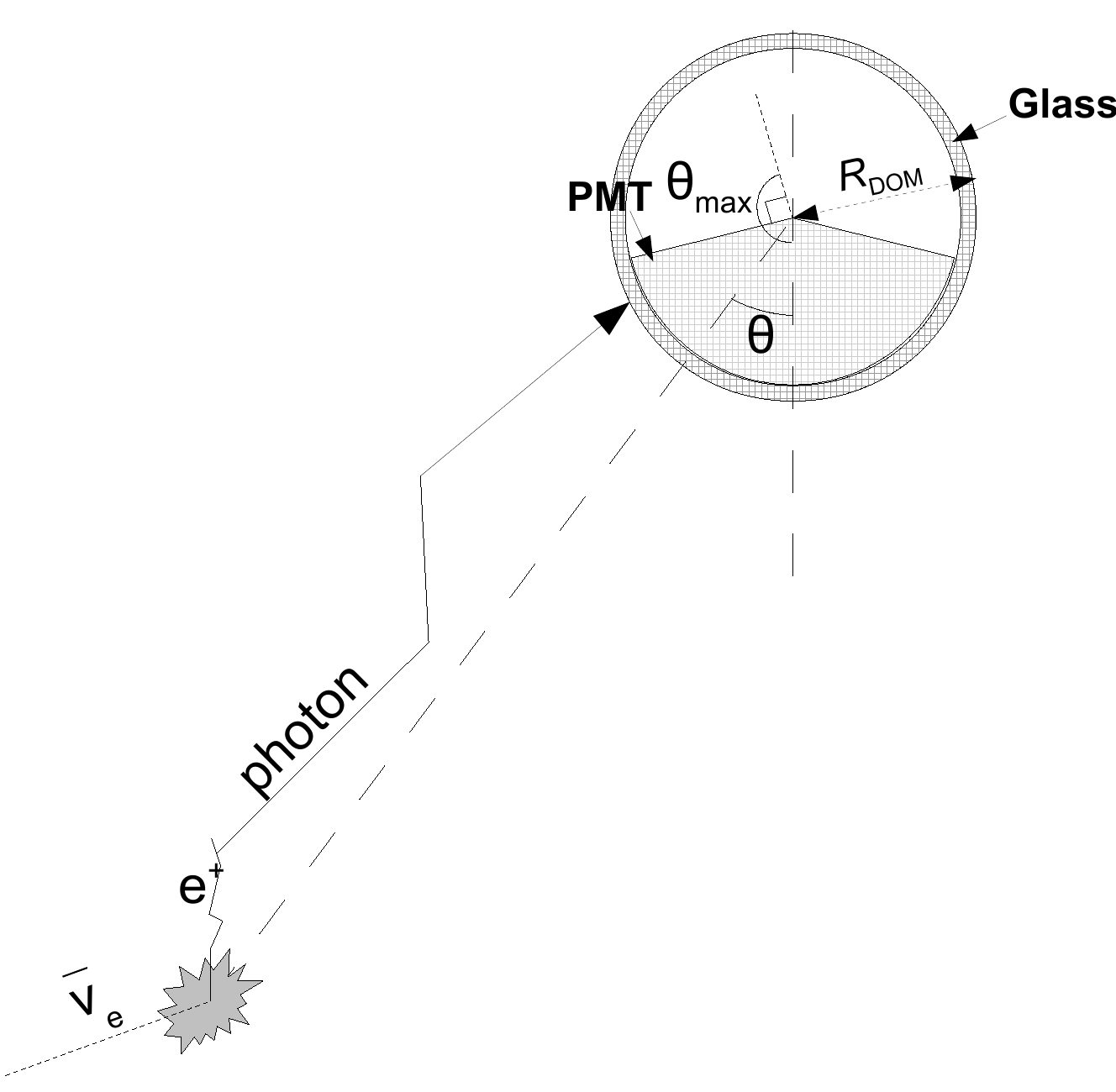}
\caption{Illustration of the sensor geometry and definition of
  $\theta$ and $\theta_{\mathrm{max}}$ . \label{fig:schematic}}
\end{figure}

\vskip3mm
{\noindent\bf Sensor geometry}
\vskip1mm
\noindent The geometry of the optical modules for the
three detector configurations IC, DC and DDC is illustrated in
Fig.~\ref{fig:schematic}. The modules~\cite{Abbasi:2010vc} are
composed of a $0.5\,$inch enveloping glass sphere (to withstand
the pressures at the large deployment depths, in particular during the
ice refreezing process) with an external radius $R\equiv
R_{\mathrm{OM}}=13\,$inch, enclosing a photomultiplier R7081-02 made
by Hamamatsu Photonics~\cite{hamamatsu-i3dom}. The glass and contact
gel between the PMT and the glass introduce a $350\,$nm lower cutoff.

In all configurations except DDC$_{4\pi}$, the photocathode sensitive
surface of the sensor covers a fraction of the sphere slightly smaller
than the half sphere. The resulting cutoff angle on the sensitive
surface is given by $\theta_{\mathrm{max}}-\pi/2$, where
$\theta_{\mathrm{max}}$ is the limit on the incoming direction of
Cherenkov photons at which they can intersect the sensitive surface of
the module. Given the specifications of the IceCube photomultiplier, we
can estimate $\pi-\theta_{\mathrm{max}}\approx 25^\circ$.

Viewed under an angle $\theta$, the projected area of the sensitive
surface is 
\[A\equiv A_{\mathrm{OM}}(\theta) = \pi R R(\theta),\]
where
\[R(\theta) = R
\frac{1-\cos{(\theta_{\mathrm{max}}-\theta)}}{1-\cos{\theta_{\mathrm{max}}}}\,H(\theta_{\mathrm{max}}-\theta)\] 
for a module sphere looking down and with a photocathode sensitive
area cutoff angle $\theta_{\mathrm{max}}-\pi/2$, as illustrated on
Fig.~\ref{fig:schematic}.

\vskip3mm
{\noindent\bf Sensor photon detection efficiency}
\vskip1mm
\noindent 
The photon detection efficiency (PDE) of the sensitive surface
$\chi(\lambda)$ has been measured
somewhere~\cite{japanPublicGoldenPMTpages}, it combines the PMT Q.E.,
the transparency of the surrounding glass and the contact gel which
couples the glass to the PMT. Averaged over a $\lambda^{-2}$ Cherenkov
spectrum between $300$ and $650\,$nm, and accounting for an amplitude
trigger threshold keeping $\epsilon_{\rm{thr}}=\,$90\% of the p.e.'s,
it yields an integrated PDE
\[\langle\chi\rangle=\frac{\epsilon_{\rm{thr}}}{\int
  d\lambda/\lambda^2} \int \chi(\lambda)/ \lambda^2 \,d\lambda  = 7.3
\%\]
for the regular IC sensors. The high quantum efficiency
sensors deployed in DC are assumed to have a PDE 30\% higher
\cite{Koskinen:IceCube}. The same 
assumption is made for the DDC and DDC$_{4\pi}$ configurations. In
contrast to previous works~\cite{halzen-sn}, $\langle\chi\rangle$ is
significantly lower, due to a regular optical module peak PDE of about
14\% at $420\,$nm.  

Although $\langle\chi\rangle$ is a fair approximation for our
analytical estimates presented in the next section, a toy Monte Carlo
simulation should implement the complete wavelength dependence of
$\chi(\lambda)$ given that the propagation of photons in the ice
medium is described in terms of wavelength-dependent scattering and
absorption lengths $\lambda_{\rm{scatt}}$ and
$\lambda_{\rm{abs}}$~\cite{icecube-icepaper}. This dependence in the
propagation cannot be disentangled from that of the PDE. 

\vskip3mm
{\noindent\bf Sources of background noise}
\vskip1mm
\noindent 
The following types of background may limit the applicability of the
coincident hit detection methodology:
\setcounter{paragraph}{0}
\vskip1mm\paragraph{Cosmic-ray induced muons.} A fraction will travel
down to the detector and produce coincident hits. We assume that most
of these events are muons dying in the upper layers of the detector:
they barely enter the fiducial volume and produce light on a
very limited number of external sensors. We assume that
this event class can be adequately rejected. Some higher energy muons
however penetrate deeper into the
detector and although most of them can be vetoed given their specific
topology, some may leave a signature well inside the instrumented volume
mimicking the coincidence signature under investigation. We
would expect the resulting rate to be minimal and thus neglect it, but this 
specific topology should still be further studied separately by means of the 
experimental data, to properly substantiate this statement.

\vskip1mm\paragraph{Low $E$ atmospheric neutrino-induced muons or
  electronic showers.} This background is constituted by low energy
neutrinos in the $10\,\mathrm{MeV} - 1\,\mathrm{GeV}$ range
\cite{solarNu,Honda:atmosnu,Barr:atmosnu}.
 The interaction rate is very limited, well below $1\,$Hz,
in a Gton ice volume.

\vskip1mm\paragraph{Sensor correlated noises at $\mu$s and ms time
scales, due respectively to photomultiplier afterpulse and radioactive
decay chains}\cite{pmtTechnical}. The afterpulse components of the
dark noise come with well defined~\cite{R7081-afterpulse-study,
Abbasi:2010vc} delays after the original photo-induced signal, larger than 
the time windows used below for coincidental hit detection within a
single sensor, so that (1) it does not pollute and therefore impede
the coincidental hit detection method for this specific channel, (2)
it can be filtered out by means of an offline analysis, leading to an
effective noise rate significantly smaller than measured (dark count
reduced from $580$ Hz to $r\equiv r^{10}=440\,$Hz~\cite{Abbasi:2010vc,
  icecube-sn}). This reduced dark count rate value is thus
considered to calculate the probability of the noise. Note that the
deadtime introduced by this afterpulse filter is negligible
($\lesssim$ 0.5\%).

The effect of ms timescales of correlated noise is more complex and it
is not clear whether it would affect the potential of the new
methodology when applied to specific channels with multiple hits
in a single sensor.

%
\vskip1mm
We define the hit times $\{t_i\}_i$, a hit time window by its width
$(\delta t)_{ij}$ and an arbitrary time shift $t_{ij}$. Hits $i$ and
$j$ belong to the time window if $|(t_j-t_i)-t_{ij}|<(\delta
t)_{ij}/2$. We restrict the study to several qualitatively complementary NN detection modes,
\begin{itemize}
\item single sensor hit: mode $10$ or "1+0";
\item two photons detected in the same sensor: mode $20$ or "2+0";
\item a single hit detected in two NN sensor: mode $11$ or "1+1";
\item three photons detected at times $t_1,\,t_2$ (in the same sensor,
  $t_1<t_2$) and $t_3$  (in the sensor above): mode $21$ or "2+1".
\end{itemize}

The noise rates of two hit coincidences $r^{20}$ and $r^{11}$
given by
\[r^{11} = r^2\, (\delta t)_{13} \mathrm{\ and\ } r^{20} = 2 r^2\, (\delta t)_{12}\]
(the latter factor two because hits at a single module are
indistinguishable). The coincidental noise rate with two hits in the
lower sensor and a single hit in its upper neighbor is 
\[r^{21}=2 r^3\, (\delta t)_{12}\, (\delta t)_{13}.\]
Note that in the IceCube detector, only the lower hemisphere of the
optical module is sensitive and therefore, a hit in the upper sensor
coming after a hit in the lower sensor is strongly preferred (i.e. an
optimal time window will have $t_{13}>0$). This is not true for some
of the futuristic detector configurations which have a uniform
sensitive surface: in this case, we will choose $t_{13}=0$.

The detector coincidental noise rate scales linearly with the number
of sensors (the number of independent pairs is almost equivalent to
the number of sensors, provided that the number of sensor along a
string is $\gg1$). 

IceCube has a system that records local coincidences within a
$2\,\mu$s ($\pm1\,\mu$s) time window up to the next-to-nearest
neighbor (there are four such neighbors) \cite{Abbasi:2008ym}. The rate of such
coincidences, called {\it hard local coincidences} (HLC), should
be therefore of the order of $r^{\rm{HLC}}_{1\,\mu\rm{s}} = 4\times
2\times r_{1\,\mu\rm{s}} \approx 1.55\,$Hz per optical module. For our
purpose, we consider the rate of coincidental events within a much
shorter coincidence time window with $\Delta t=30-150\,$ns involving
one neighboring sensor corresponding to a coincidental noise rate of
$r_{30-150\,\rm{ns}} \approx 0.006-0.03\,$Hz per sensor. 

A SN signal becomes significant when the number of coincidental events
integrated over the typical duration of the SN neutrino emission
(about 3 seconds) takes values larger than the fluctuations of the
noise rate $r^{ij}$, corresponding to the various event classes.

\subsection{Supernova neutrino emission models}
We have chosen to study several flux models of rather modest
luminosity. Our main benchmark flux model is the "Sf"
model (also referred to as "Garching" model), an 8.8 solar mass electron capture supernova, discussed
in~\cite{garching} and which may represent a large fraction of all
supernovae.

Furthermore, in this paper we will only consider the predicted
anti-electron neutrino flux, focusing solely on the inverse beta decay
(IBD) reaction in ice, {\it i.e.} $\bar{\nu}_{\mathrm{e}} + \mathrm{p}
\rightarrow \mathrm{n} + \mathrm{e}^{+}$. For one, in water or ice,
which are natural media for high energy neutrino telescopes, the
(anti-)electron neutrino interaction with 
oxygen has a much higher energy threshold and a quite significantly
smaller CC cross section at the energies of interest
\cite{Kolbe:oxygen}. This  largely makes this
channel subdominant for the interaction rate in the detector
compared to the IBD reaction yield. Moreover, a large fraction of the
neutrino energy is absorbed into binding energy, reducing the
Cherenkov light yield and dramatically suppressing the potential of
this reaction channel given that our methodology relies on the coincident detection of
Cherenkov photons in neighboring sensors. We also neglect the
elastic interaction channel $\bar{\nu}_{\mathrm{e}} + \mathrm{e}^{-} \rightarrow
\bar{\nu}_{\mathrm{e}} + \mathrm{e}^{-}$ which has an even lower cross section
than the oxygen channel \cite{Marciano:elastic}.
 
The anti-electron neutrino flux luminosity is given by
\[ \frac{{\rm{d}}\tilde{\Phi}_{\bar{\nu}_{\rm{e}}}}{{\rm{d}}E_\nu}(E_\nu,\,d)
= \frac{1}{4\pi d^2} \int_0^{\tilde t} \frac{L(t)}
{\langle E_\nu \rangle(t)}
f_{\alpha(t),\langle E_\nu \rangle(t)} {\mathrm{d}}t 
\]
where
\[f_{\alpha,\langle E_\nu \rangle}(E_\nu) = \frac{(1+\alpha)^{1+\alpha}}{\langle E_\nu \rangle \Gamma(1+\alpha)}
\left(\frac{E_\nu}{\langle E_\nu \rangle}\right)^\alpha {\mathrm{e}}^{-(1+\alpha)\frac{E_\nu}{\langle E_\nu \rangle}}\label{eq:fAlphaAvEnu}
\]
and with the time-dependent model parameters luminosity $L(t)$,
average neutrino energy $\langle E_\nu \rangle(t)$ and pinch parameter
$\alpha(t)$. Their time evolution can be found in~\cite{garching}. It
is interesting to note, as illustrated in Fig.~\ref{fig:
  timeIntPositronspectrum}, that the parameters $\alpha$ and $\langle
E \rangle$ can be very well fitted with Eq.~\ref{eq:fAlphaAvEnu} in order to match the integrated
neutrino flux and positron density up to a certain time. 
\begin{figure}[h]
  \centering
  \includegraphics*[width=0.5\textwidth,angle=0,clip,trim=3cm 21cm 9cm 2cm]{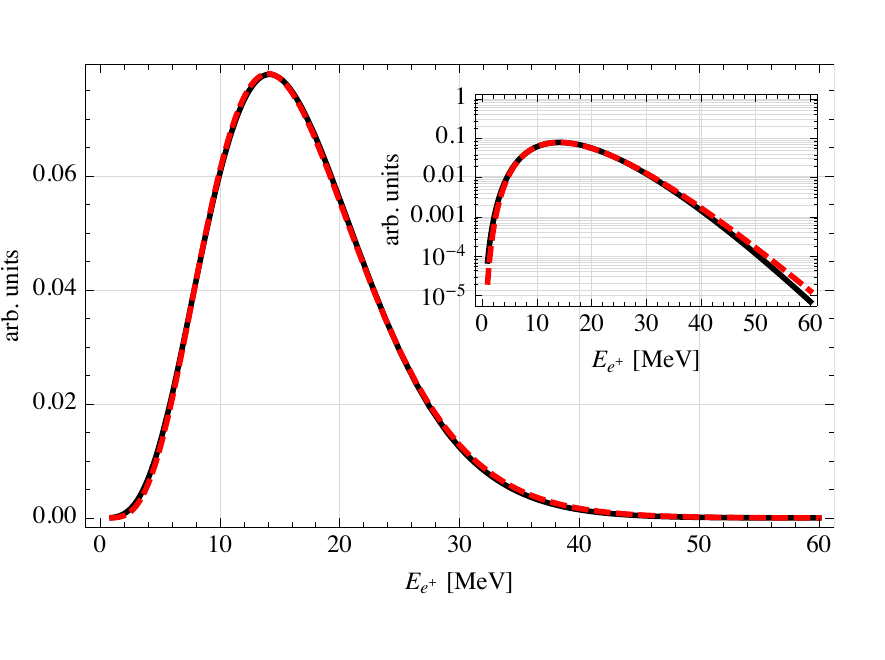}
  \caption{Positron spectrum, integrated up to $\tilde t$. Dashed line is the result of fitting $\alpha=4.77$ and $\langle E_{\mathrm{e}} \rangle=16.97\,$MeV.\label{fig: timeIntPositronspectrum}}
\end{figure}

The neutrino flux integration time $\tilde{t}$, which closely
optimizes the potential in terms of distance reach of the coincident
hit method is derived by maximizing the signal to the square root of noise
ratio, assuming an exponential luminosity decrease, in good
qualitative agreement $\forall t \gtrsim 1\,$s, with $t_0\approx
2.35\,$s, of the luminosity evolution in the Garching model (a usual
luminosity decrease in $(1+t/t_0)^{-\beta}$ does not fit well).

The argument could be more sophisticated, e.g.\ by using a refined
likelihood temporal analysis.

The integrated signal (the supernova neutrinos producing an event) up
to time $\tilde t$ is $S(\tilde t)= \int_0^{\tilde{t}} s_0 {\rm
  e}^{-t/t_0} {\rm{d}}t =s_0 t_0 (1-{\rm e}^{-{\tilde{t}}/t_0})$. The
detector noise is assumed gaussian with count fluctuation of a
complete detector configuration $\sqrt{\tilde{\mu}}=\sqrt{N_{\rm{OM}}
  r \tilde{t}}$ within an integration time $\tilde{t}$, where $r$ is
the noise rate of a specific mode. 
$\tilde{t}$ is determined
\[
\frac{{\rm{d}}}{{\rm{d}}\tilde{t}} \frac{S(\tilde{t})}{\sqrt{\tilde{\mu}}} = 0
\]
leading to 
\begin{equation}
  \tilde t= \frac{t_0}{2} ({\rm e}^{{\tilde t}/t_0}-1).
\label{eq:transEqu}
\end{equation}
We extract $\tilde t = 2.95\,$s for $t_0=2.35\,$s and we note that 
$\tilde{t}$ is independent of $s_0$ (proportional to the
luminosity), of the noise rate $r$ and of the significance of the
observation.

The luminosity at early times is larger than in this description, so
that this integration time $\tilde t$ constitutes a conservative upper
bound regarding the potential of the method: if more precisely
assessed, $\tilde{t}$ and thus the integrated noise would be reduced.

The integrated anti-electron neutrino luminosity within this time-span
$\tilde{t}$ is $\tilde{L} = 2.13\cdot10^{52}\,$erg, and the flux
$\tilde{\Phi}_{\bar{\nu}_{\rm{e}}}=0.89\cdot10^{15}\,{\rm{m}}^{-2}$. The
average neutrino energy is $\langle
\tilde{E}_{\bar{\nu}_{\rm{e}}} \rangle = 12.5\,$MeV. $\tilde{L}$ is
about three quarter of the total luminosity. It is important to note,
when comparing the limiting distance up to which a SN can be
discovered, that the total luminosity is about 56\% of the one considered in~\cite{KistlerAndBeacom}.

We have also calculated the outcome of the novel detection methodology
for neutrino spectra which obey a Fermi-Dirac distribution,
\[f_{\eta, T}(E_\nu)=N(\eta, T)
\, \frac{(E_\nu/  T)^2}
{1+  {\mathrm{e}}^{E_\nu/T - \eta}
}
\]
where $N(\eta, T)=-(2T Li_3(-{\mathrm{e}}^\eta))^{-1}$. We have normalized the amount of
energy released over $\tilde{t}$ to be equal to the Sf model, for two
temperature, $T=5\,$MeV and $T=6.5\,$MeV. Although the average neutrino
energies, $\langle E_{\bar{\nu}_{\rm{e}}} \rangle_{T=5\,{\rm{MeV}}} =
15.8\,$MeV and $\langle E_{\bar{\nu}_{\rm{e}}}
\rangle_{T=6.5\,{\rm{MeV}}} = 20.5\,$MeV, are larger resulting in
smaller flux than in the Sf model, the interaction density is
increased by more than 50\% due the IBD cross section increase turning
into a considerable increase in the coincidental hit rate (by a factor
$>2$) as will be seen.

\subsection{Discovery potential and sensitivity}
We assess the discovery and learning potential by requesting the
signal over square root of noise ratio $s=5$ and $s=1$
for the C.L. in $\sigma$ unit, respectively, using the Poissonian distribution
$p_\mu$: given a mean number of expected coincidental noise events
$\tilde{\mu}$, a minimum number of $\ge m$ events is required to reach
significance $\ge s$, 
\[
\min_m \left\lbrace\right. m\,\left|\right.\, p_{\tilde{\mu}}(0)+\sum_{k=1}^{m}p_{\tilde{\mu}}(k)\ge{\rm{erf}}(\frac{s}{\sqrt{2}}) \left.\right\rbrace.
\] 
The distance reach is therefore given by 
\[d^{s}_{10\,\rm{kpc}}= \sqrt{n_s / (m - \tilde{\mu})},\] where
$d^{s}_{10\,\rm{kpc}}$ is the distance in 10 kpc units and $n_s$ is
the number of signal events from a SN explosion at 10 kpc.




An advantage of the detection method relying on coincidental hits
arises from a more gaussian distributed coincidental noise rate (while
the single hit noise rate distribution has non gaussian tails,
requesting coincidences dilutes them).
The false alarm rate (assuming gaussianity) for $s=5$ is
roughly $\tilde{t}/{\rm{erfc}}(5/\sqrt{2})/2=3\,\rm{yr}^{-1}$.

$s=1$ is a requirement too weak
and not adequate for IceCube to trigger as a standalone detector, but
where IceCube may help establishing stronger constraints by enhancing
the quality of an observation:
it sets the distance scale for IceCube
to contribute in supplementing information to observations made by other
instruments detecting e.g. a neutrino or an optical signal. In the
latter case, of course, IceCube would have to buffer the relevant
information because of the few hours delay between the neutrino and the
optical signal.



\section{Estimate of single and coincident hit detection rate}
The positron effective detection volume 
as a function of the neutrino energy is the product of the number of
target protons in the interaction volume and of the fraction of
uniformly distributed positrons leading to a detection. 
It is an important and well defined concept for our analysis because
the positron tracks are nearly point-like with a path length of about
10 cm in average, a small distance compared to the mean distance
between a positron and a sensor in which it leaves a signature.
%
%
This enables the estimation of the response of the detector to a given
flux of low energy neutrinos.

The positron effective volume for coincidence of order $k$ at a single
module is denoted $V_{\mathrm{eff}}^k(E)$, where $k=1,2$ for the
detection of a single hit resp.\ of two coincident hits. We write
$V_{\mathrm{eff}}^{kk'}(E)$ for a detection of order $k$ in one sensor and of
order $k'$ at the NN.

The effective volume depends only slightly on the incoming
neutrino direction $\vartheta$, due to the proximity of the average
neutrino energy with the energy at which the anisotropy in the positron
emission (transition between backward and forward scattering) from the
IBD interaction vanishes. Note however that selecting 
events in the high energy tail of the neutrino
emission, where forward scattering is favored, may grant access to
the direction of a nearby supernova. This can be done to some extent by
means of the coincident hit time analysis.

The current supernova detection strategy with IceCube relies only on
$V_{\mathrm{eff}}^1(E)$. We analyze here the power of the new strategy
by relying more generally on $V_{\mathrm{eff}}^{kk'}(E)$. We emphasize
as well the importance of certain ratios, such as
$V_{\mathrm{eff}}^2(E)/V_{\mathrm{eff}}^1(E)$ or
$V_{\mathrm{eff}}^{11}(E)/V_{\mathrm{eff}}^1(E)$, which open new
perspectives and grant access to new observables, e.g. the average
positron energy.

For a given supernova neutrino emission model ${\mathcal{M}}$,
time-integrated up to time $\tilde{t}$, the effective volume
definition can be extended to embed the spectral shape of the positron
energy distribution
${\mathrm{d}}\tilde{\rho}_{{\mathrm{e^+}}}^{\mathcal{M}} /
{\mathrm{d}}E$,
\[\tilde{V}_{\mathrm{eff}}^{{\mathcal{M}},\,kk'} =
\frac{1}{\tilde{\rho}_{{\mathrm{e^+}}}^{\mathcal{M}}(d)} \int
\frac{{\mathrm{d}}\tilde{\rho}^{\mathcal{M}}_{{\mathrm{e^+}}}}{{\mathrm{d}}E}(E,d)
V_{\mathrm{eff}}^{kk'}(E) {\mathrm{d}}E,
\label{veffmodel}
\]
where
\[ \tilde{\rho}_{{\mathrm{e^+}}}^{\mathcal{M}}(d)
=
\int \frac{{\mathrm{d}}\tilde{\rho}^{\mathcal{M}}_{{\mathrm{e^+}}}}{{\mathrm{d}}E} (E,d)
{\mathrm{d}}E
\]
and which follows from the definition \[\tilde{N}^{{\mathcal{M}},\,kk'}(d) =  \tilde{\rho}^{\mathcal{M}}_{{\mathrm{e^+}}}(d) \tilde{V}^{{\mathcal{M}},\,kk'}_{\rm{eff}}.\label{eq:NVR}\]

From now on, we drop the ${\mathcal{M}}$ or replace it by one of the specific models. The presence of the tilde denoting time-integration necessarily refers to a model and confusion in the notation is avoided.

The differential positron density ${\mathrm{d}}\rho_{\mathrm{e^+}} /
{\mathrm{d}}E_{\rm{e}}$ is extracted from the incoming neutrino flux
${\mathrm{d}}\Phi_\nu / {\mathrm{d}}E_\nu$ by folding it with the
differential IBD cross section
${\rm{d}}\sigma/{{\rm{d}}}E_{\rm{e}^{+}}$  given by Eq. (10)
in~\cite{vissani-strumia} (accurate up to NNLO in $\epsilon$),
\[ \frac{{\mathrm{d}}\tilde{\rho}_{{\rm{e}^{+}}}}
{{\mathrm{d}}E_{\rm{e}}}(E_{\rm{e}},\,d)=
\rho_{\rm{p}}\int\frac{{\rm{d}}\sigma}{{\rm{d}}E_{\rm{e}}}(E_\nu,E_{\rm{e}})\,
\frac{{\mathrm{d}}\tilde{\Phi}_{\bar\nu_{\rm{e}}}}{{\mathrm{d}}E_\nu}(E_\nu,\,d)\,
 {\mathrm{d}}E_\nu \]
and the total number density of positron interactions (same luminosity
for all three models),
\[ \tilde{\rho}_{{\rm{e}^{+}}} (10\,{\mathrm{kpc}})=
\left\{ \begin{array}{ll}0.074\,{\mathrm{m}}^{-3},& \rm{Sf}, \\
    0.096\,{\mathrm{m}}^{-3}, & {T_{\rm{FD}} = 5\,{\rm{MeV}}}, \\
    0.12\,{\mathrm{m}}^{-3}, & {T_{\rm{FD}} =
      6.5\,{\rm{MeV}}} \end{array} \right. 
\]
%
%
%
The proton target density is given by $\rho_{\rm{p}}=2\,\rho_{\mathrm{ice}}N_a/M_{\mathrm{ice}}$,
where $\rho_{\mathrm{ice}} = 0.91\,\mathrm{g/cm}^3$, $M_{\mathrm{ice}}
= 18.015\,\mathrm{g/mol}$, and $N_a$ is the Avogadro constant.

The number of expected events belonging to the various classes
$\tilde{N}_{\mathrm{e}^{+}}^{kk'}$ (see Eq.~\ref{eq:NVR}), obtained by
integration of
\[
\frac{{\rm{d}}\tilde{N}_{\mathrm{e}^{+}}^{kk'}}{{\mathrm{d}}E_{\rm{e}}}
=   V_{\mathrm{eff}}^{kk'}(E_{\rm{e}})
\frac{{\mathrm{d}}\tilde{\rho}_{{\rm{e}^{+}}}}
{{\mathrm{d}}E_{\rm{e}}}(E_{\rm{e}},\,d),  \label{eq_positronSpectrum}
\]
also defines the relevant observables
$\tilde{r}_{kk'}^{jj'}=\tilde{N}_{\mathrm{e}^{+}}^{kk'}/\tilde{N}_{\mathrm{e}^{+}}^{jj'}$,
some of them being particularly sensitive to the average energy of the
neutrino spectrum.








\subsection{Cherenkov emission}
In order to proceed with the effective volume estimates, we need to
approximate the average number of detectable Cherenkov photons emitted
along the track of a positron with energy $E$, which could naively be
done by combining the Franck-Tamm \cite{franck-tamm} ($\beta=1$), the
average photomultiplier Q.E. and positron range formulas,
\begin{equation}
N(E) = 2\pi\alpha \, R(E)  \,  \left(1 -
  \frac{1}{\beta^2n_{\mathrm{ice}}^2}\right)  \int \frac{\chi(\lambda)
  d\lambda}{\lambda^2} \label{eq:ne}\end{equation}
where the integrals for average Q.E and Franck-Tamm formula run
between 300 and 650 nm. $R(E)$ is the positron track length above
Cherenkov emission threshold and given
by~\cite{wilson-range,pal-stopping_power,pal-range}:
\[R(E)=\ln{2}\, \ln{(1+E/(E_{\mathrm{c}} \ln{2}))}\,
X_0^{\mathrm{water}} / \rho_{\mathrm{ice}},\label{eq:range}\]
where $X_0^{\mathrm{water}}=36.08\,{\mathrm{g/cm^2}}$,
$\rho_{\mathrm{ice}}=0.92\,{\mathrm{g/cm^2}}$ and
$E_c=610\,{\mathrm{MeV}} / (Z+1.24)=72.1\,$MeV~\cite{pdg}.
\begin{figure}[h]
\centering
 \includegraphics*[width=0.45\textwidth,angle=0,clip]{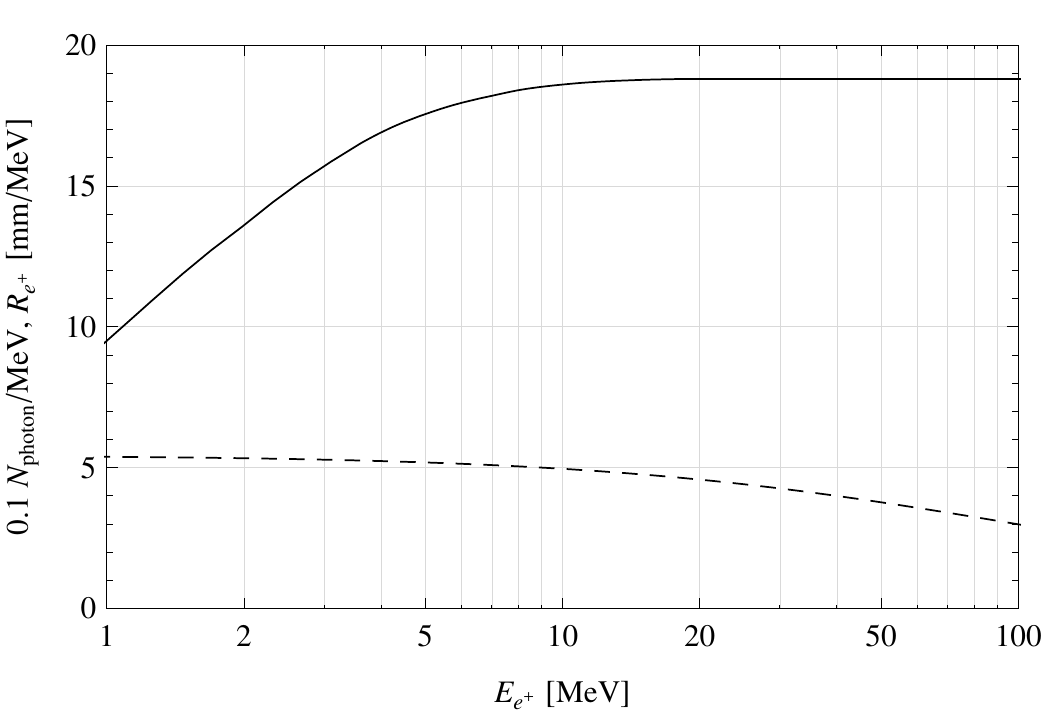}
\caption{Number of emitted photon / 10 / MeV by a positron and its
  byproducts between 300 and 650 nm (Geant4) (solid line) and positron
  range parametrization above Cherenkov emission threshold in mm/MeV
  (dashed line) w.r.t.\ the positron energy.}\label{fig:rangeAndNE-positron}
\end{figure}

E.g.\ at $E=10\,$MeV, $R(E)\approx 5.5\,$cm and the average number of
detectable photons (IC) by a positron is $N\approx
135$. However, Eq.~\ref{eq:ne} does not take
into account Cherenkov photons which are emitted by secondary
particles,e.g.\ knock-off delta electron and positron
annihilation. Results from a Geant4 simulation show that the total number of
emitted photons remains almost constant
(c.f.~Fig.~\ref{fig:rangeAndNE-positron}). In the following we will
use the Geant4 photon yield results for the calculations.





\subsection{Effective detection volume}
The fraction $\epsilon(r,\theta)$ of the $4\pi\,$sr solid angle
occupied by the sensitive surface of the module, viewed under angle
$\theta$ and at distance $r$ is
\[ \epsilon(r,\theta) = \frac{1}{2}\left(1-\frac{1}{\sqrt{1+R
      R(\theta)/r^2}}\right).\]
(we have neglected the spherical shape of the sensor but this remains
an excellent approximation).

When $r \gg R$, this expression 
%
reduces to
%
%
\[
\epsilon(r,\theta) =
\frac{A_{\rm{DOM}}(\theta)}{4\pi r^2}.
\]



The expectation number of detected photons is
\[\mu(E,r,\theta)=N(E)\epsilon(r,\theta)\]

We have now all the necessary ingredients to express the coincident hit effective volumes,




\[
V_{eff}^k(E_{\rm{e}}) = 2\pi\, \int_{R}^{\infty} r^2 {\mathrm{d}}r
\int_{-1}^{1}{\mathrm{d}}\cos{\theta} f(k,\,r,\,\theta,\,E_{\rm{e}}).
\label{veffk}
\]
\[
f(k,\,r,\,\theta,\,E_{\rm{e}}) = \frac{1}{4} \left( 1-\sum_{l=0}^{k-1}
  p_{\mu(r,\,\theta,\,E_{\rm{e}})}(l) \right) \label{eq:fMu}
\]
is the density of probability of recording $\ge k$ hits from a
positron with energy $E_{\rm{e}}$, where
$p_{\mu(r,\,\theta,\,E_{\rm{e}})}(k)$ is a Poisson distribution with
mean value 
\[
\mu(r,\,\theta,\,E_{\rm{e}})=4 N(E_{\rm{e}}) \epsilon(r,\,\theta)
\exp{(-r/\lambda_{\rm{abs}})} \label{eq:muR}
\]

For the effective volume for a sensor with $k$ hits and NN
sensor with $k'$ hits, we can write the more generic expression,

\begin{eqnarray}
  V_{eff}^{kk'}(E) &=&  2\pi\, \int_{R}^{\infty} r^2 {\mathrm{d}}r   \int_{-1}^{1}{\mathrm{d}}\cos{\theta} \nonumber \\ &\times& 4  f(k,\,r,\,\theta,\,E_{\rm{e}})  f(k',\,r',\,\theta',\,E_{\rm{e}})
  \label{veffkk} 
\end{eqnarray}
where $r'=\sqrt{r^2+L^2+2Lr\cos{\theta}}$ and $\cos{\theta'}=(L+r\cos{\theta})/r'$. 

According to~\cite{halzen-sn} and for the purpose of this
estimate, we have made the qualitative assumption that the Cherenkov
photons are emitted in correlated directions, roughly over $\pi\,$sr,
or a quarter of the sphere, the approximate size of the Cherenkov
cone. We therefore effectively implement the anisotropic emission with
the inclusion of several factors $4$. 
In order to obtain simple expressions, we have neglected the spacing
between the modules which do affect the probability of two NN sensors
to be in the pool of photons. This may have a significant impact on
the effective volume. We qualitatively argue that most interactions leading to
a coincidental hit detection occur slightly below the lower sensor, i.e. $\theta$
is in general close to $\theta'$. However, this argument does not hold for 4$\pi$
sensitive modules. In this case, the expressions derived here are
quite approximative.

The correlated probability of photon detection is justified at short
distances from the sensor, as the photons are not yet diffused. This
correlation gradually vanishes at larger distances and fully
disappears at an effective scattering length distance. Fortunately, our
assumption is likely quite accurate for NN coincidental hit
detection from low energy events, because the detection
probability of $>1$ photon is sizable only at short distances and the
photon directions are not yet decorrelated (i.e. the fraction of
contributed events beyond a relatively short distance is negligible,
due to a negligible fraction of the subtended areas of the
sensors).

The validity of the effective volume formula for single hits with the 
introduced anisotropic emission holds for arbitrary distance: 
although large values of the integration variable $r$ 
(i.e at distances where the Cherenkov photons and the initial positron 
directions are decorrelated) 
contribute significantly to the integral, 
there is an equivalence  at large distances between 
isotropic emission - uniform low photon expectation value and anisotropic emission - higher photon expectation value within the $\pi\,$sr cone.
Stated differently, the equivalence at large distances (i.e. $\mu$ small)  $f(\mu) \approx 4f(\mu/4)$ (see the expression for $f(\mu)$, Eq.~\ref{eq:fMu}).


In Eq.~\ref{eq:muR}, we have introduced a damping factor related to
the absorption length. One may question whether we should consider
instead the attenuation length as appropriate. This can be understood
from first principles (phase space conservation): photons do not
propagate to such large distances as they scatter and will only reach
an average distance of about an attenuation length from the
sensor. However, the subtended sensor area for the photon detection
decreases with distance and the large absorption length considered
compensates for the shorter effective distance reached by an average
photon but seeing a larger subtended photo-sensor area.

\begin{figure}[h]
\centering
 \includegraphics*[width=0.45\textwidth,angle=0,clip,trim=0mm 0 0 0]{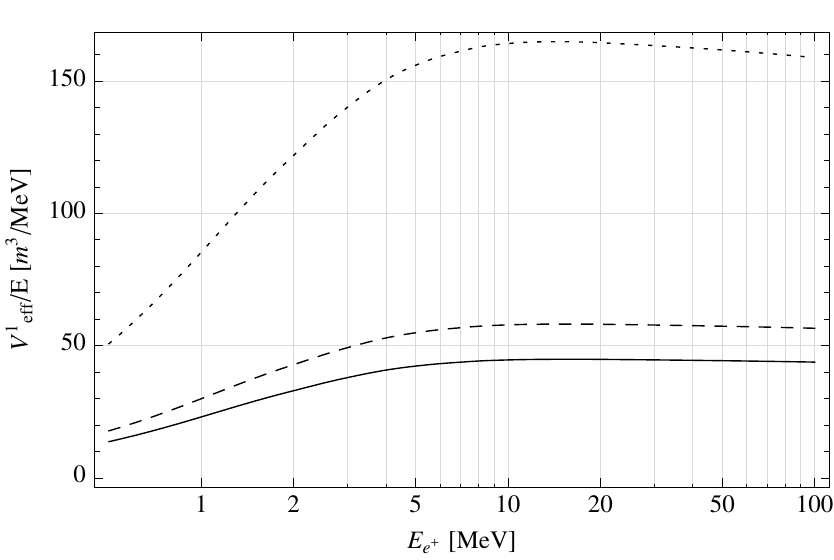}
 \includegraphics*[width=0.45\textwidth,angle=0,clip,trim=0mm 0 0 0]{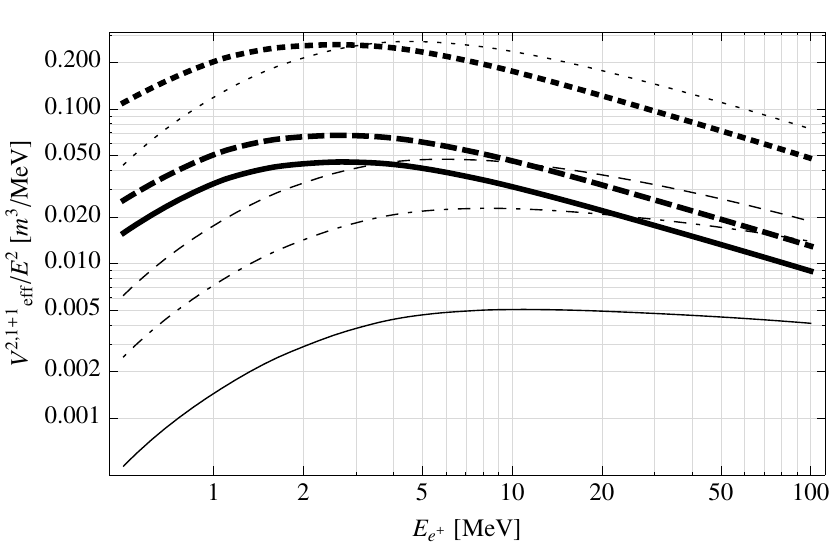}
 \includegraphics*[width=0.45\textwidth,angle=0,clip,trim=0mm 0 0 0]{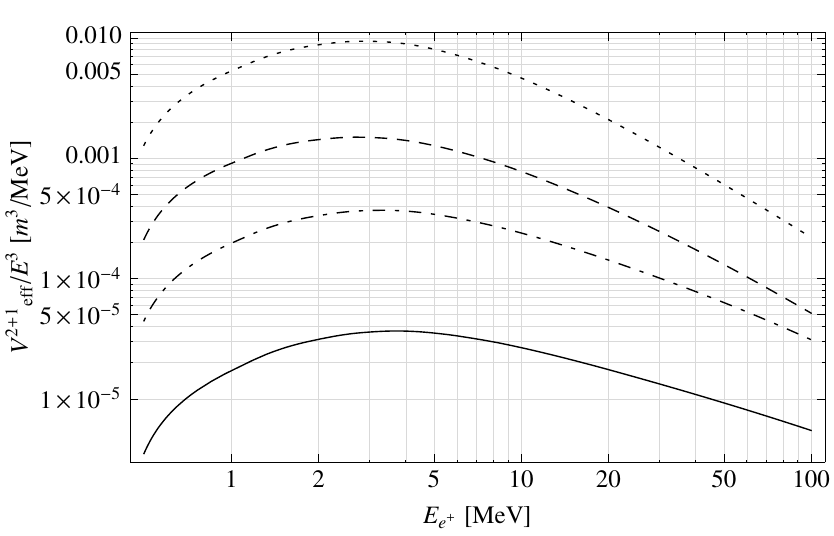}
\caption{Upper / middle /lower plots: effective volumes for single hit
  / coincidental hit / next order coincidental hit. Dotted, dashed,
  dot-dashed and solid curves resp.\ for DDC$_{4\pi}$, DDC, DC and
  IC. Middle plot, thick / thin lines for resp.\ $2+0$ and $1+1$ modes
  (note that DC and DDC results are the same for $1+0$ and $2+0$
  modes).
\label{fig:effVol}}
\end{figure}

The results for the effective volume in the various detection modes are
shown in Fig.~\ref{fig:effVol}. It is noticeable that while the ratio of
the effective volume to the positron energy for single hit detection
mode behaves rather flat and even eventually decreases at high energy,
it continually increases for coincidental hit detection modes. The
rate of the increase is fastest for the $2+1$ mode and faster for $1+1$
than $2+0$. This behavior reflects in a corresponding increase of the
average positron energy for the chosen neutrino emission model. For
the four different topologies $1+0,\,2+0,\,1+1,\,2+1$, the relative
variation rate with energy is constant, being independent of the  specific
detector configuration, the absolute scaling of the effective
volume, which in turn depends on the quantum
efficiency, and the sensor coverage and spacing.

The effective detection volumes in the DDC configuration are
significantly higher when the sensor is sensitive over 4$\pi$, a
factor about three for $1+0$ mode, four for $2+0$ mode and six for
$1+1$ and $2+1$ modes around the Garching energy $\langle \tilde{E}_{\bar{\nu}_e}\rangle$.

\begin{figure}[h]
  \centering
  \includegraphics*[width=0.45\textwidth,angle=0,clip,trim=0 1.5cm 0 1.4cm]{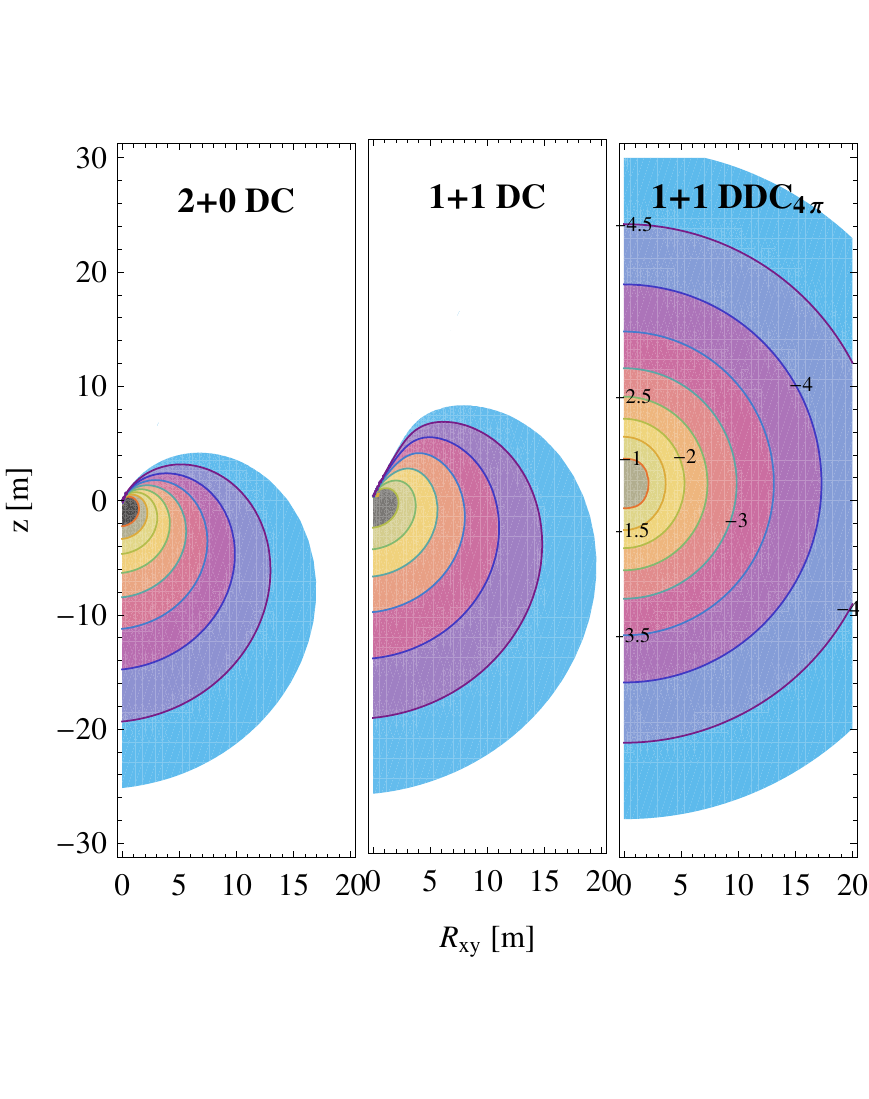}
  \caption{Detection probability as a function of the interaction for
    $1+0$ and $2+0$ modes in DC and $1+1$ mode in DDC$_{4\pi}$. The log of the equiprobability contours correspond to the values explicitly shown on the right plot. 
    \label{rhoCoinc-VS-Znu-VS-Rxy}}
\end{figure}

Fig.~\ref{rhoCoinc-VS-Znu-VS-Rxy} shows the detection probability
w.r.t.\ the IBD neutrino interaction location for DC and DDC$_{4\pi}$
configurations.

It is clear from these formulas that, while we constrain ourself to a
limited number of detection modes, it is trivial to extend the
analysis in order to obtain effective volumes for arbitrary
coincidental modes $kk'$, including those beyond our one
dimensional analysis (i.e. modes with hits on different strings which
become relevant for a hypothetical configuration with small string
spacing).

\section{Results}
\subsection{SN detection distance reach}
With the effective volumes estimated above, we now have all the
ingredients to calculate the number of single and coincidental
hit detections. In Table~\ref{table-SF}, we provide the number of
detections and the distance reach for unit signal over background
information and for a 5$\sigma$ C.L.\ detection, calculated for the
various configurations with the Sf model, and a comparison of
effective volumes in Table~\ref{table-Spectra}. Note that in the case
of the Fermi-Dirac spectrum with $T=6.5\,$MeV and the luminosity
$L_{\mathrm{BK}}$ calculated by \cite{KistlerAndBeacom}, and
large area low noise sensors, the $2+1$ detection mode reaches
${\rm{d}}_{5\sigma}^{21} = 2.9\,$Mpc and ${\rm{d}}_{5\sigma}^{11} =
7.2\,$Mpc.

\begin{table}[h]
\begin{tabular}{cc|llll}
conf. \,&\, $kk'$ & $\tilde{V}_{\mathrm{eff}}/{\rm{m}}^3$ \,&\, $n_s$ &\, $n_{\rm{bg}}$ \,&\, $d^{5\sigma,\,1\sigma}$/{\rm{kpc}}  \\
\hline
\multirow{4}{*}{IC} & 1+0 &  764 & $2.70\, 10^5$ & $6.2 \, 10^6$ & 46.5, 104 \\
				& 2+0 & 7.4 & $2.085\,10^3$ & 274 & 49.9, 108 \\
				& 1+1 & 1.7 & 450 & 411 & 20.9, 45.4 \\
				& 2+1 & 0.12 & 26.5 & 0.02 & 29.8, 52.0 \\
\hline
\multirow{4}{*}{DC} & 1+0 & 990 & $2.34\,10^4$ & $4.15\, 10^5$ & 26.9, 60.1 \\
				& 2+0 & 10.8 & 204 & 18.3 & 28.7, 109 \\
				& 1+1 & 7.0 & 122 &18.3 & 22.2, 84.0 \\
				& 2+1 & 0.98 & 13.6 & 0.001 & 36.9, 36.9 \\
\hline
\multirow{4}{*}{DDC} & 1+0 & 990 & $2.62\, 10^5$ & $4.67\,10^6$ & 49.3, 110 \\
				& 2+0 & 10.8 & $2.29\, 10^3$ & 206 & 55.9, 122 \\
				& 1+1 & 12.5 & $1.89\, 10^3$ & 103 & 59.6, 212 \\
				& 2+1 & 2.6 & 317 & 0.005 & 126, 178 \\
\hline
\multirow{4}{*}{DDC$_{4\pi}$} & 1+0 & $2.8\, 10^3$ & $7.45\, 10^5$ & $4.67\, 10^6$ & 83.0, 186 \\
				& 2+0 & 40.7 & $8.64\, 10^3$ & 206 & 109, 237 \\
				& 1+1 & 58.9 & $8.59\, 10^3$ & 103 & 127, 452 \\
				& 2+1 & 13.9 & $1.63\, 10^3$ & 0.005 & 285, 403 \\
\hline
				& 1+0 & $2.8\, 10^3$ & $7.45\, 10^5$ &$1.06\,10^6$ & 120, 269 \\
DDC$_{4\pi}$ 		& 2+0 & 40.7 & $8.64\, 10^3$ & 10.6 & 211, 791 \\
$100\,$Hz 		& 1+1 & 58.9 & $8.59\, 10^3$ & 5.31 & 242, 1116 \\
				& 2+1 & 13.9 & $1.63\, 10^3$ & 0.00005 & 403, 403 \\
\hline
DDC$_{4\pi}^{\rm{VL}}$ & 1+0 & $13.0\, 10^3$ & $8.0\, 10^6$ &$1.06\,10^6$ & 394, 880 \\
$100\,$Hz			 & 2+0 & 377 & $1.86\, 10^5$ & 10.6 & 980, 3672 \\
$T_{\rm{FD}}=5\,$MeV	& 1+1 & 582 & $1.98\, 10^5$ & 5.31 & 1160, 5354 \\
high $L_{\bar{\nu}_{\mathrm{e}}}$	& 2+1 & 171 & $46.6\, 10^3$ & 0.00005 & 2158, 2158\\
\end{tabular}
\caption{Sensor effective volume for the Garching spectrum,
  $E^{\mathrm{e}^{+}}_{\mathrm{av}}=17.1\,$MeV, various detector
  configurations and detection modes. The last entries, corresponding
  to DDC$_{4\pi}^{\rm{VL}}$, are results for an enhanced DDC
  configuration with larger modules and Q.E., for a $T=5\,$MeV
  Fermi-Dirac spectrum and a luminosity $L_{\bar{\nu}_{\mathrm{e}}}$
  normalized to $0.5\cdot 10^{52}\,$erg.\label{table-SF}}
\end{table}

\begin{figure}[h]
\centering
 \includegraphics*[width=0.5\textwidth,angle=0,clip,trim=0mm 0 0 0]{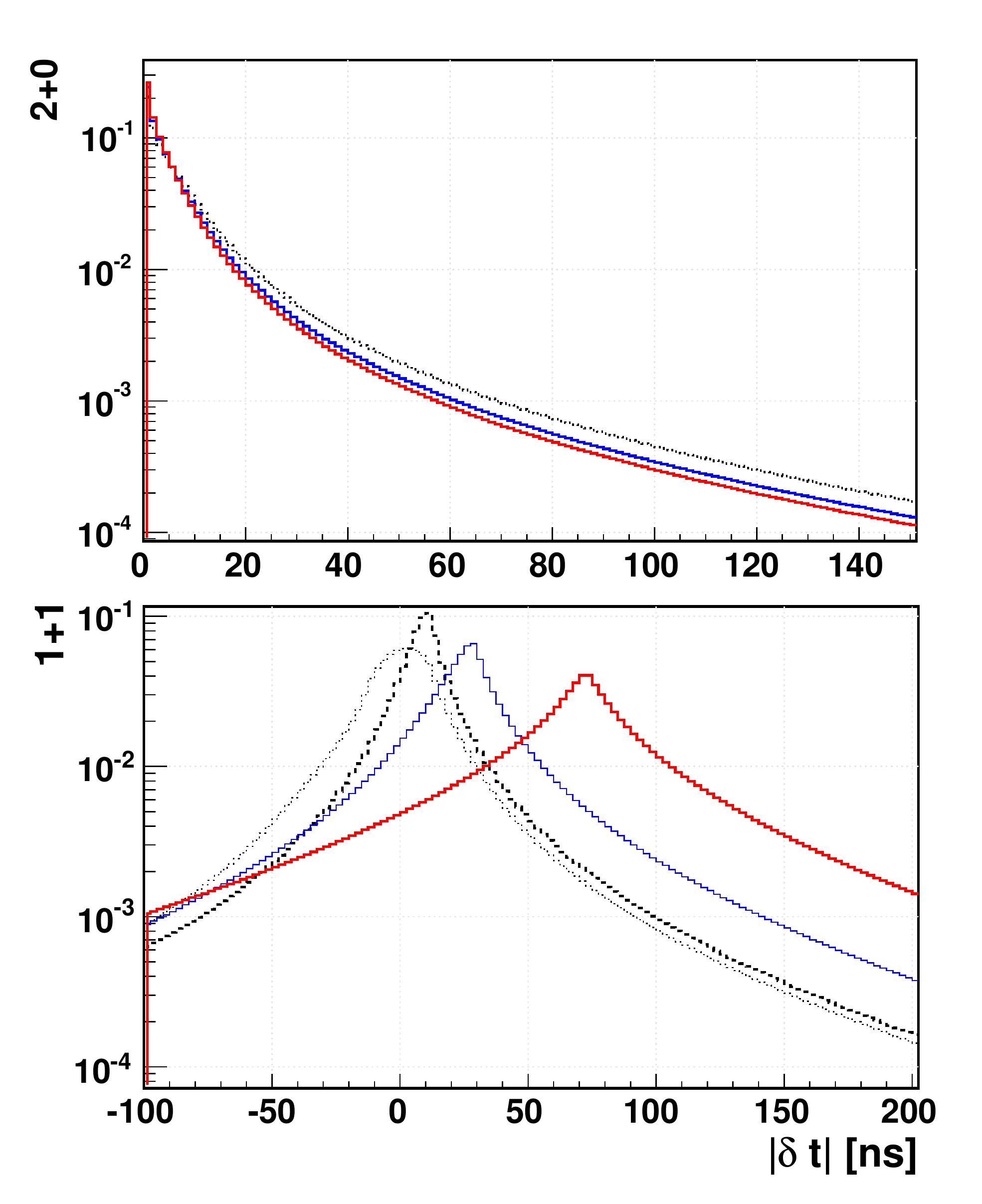}
\caption{Difference in hit arrival time. Distributions peaking from
  left to right correspond to the DDC$_{4\pi}$, DDC, DC and IC
  configurations.\label{fig:timeDiff}}
\end{figure}

The description of the distribution of hit arrival time differences
(see~\cite{ribordy-japar} and references therein) accounts for
the characteristics of the ice, which influences the photon
propagation. The shape of this distribution depends on the distance of
the sensor from the photon emission location as illustrated in
Fig.~\ref{fig:timeDiff}, which shows the arrival time difference of
hits at one sensor and its (upper) neighbor (non uniform sensor). We
notice that the shape in $2+0$ 
mode depends weakly on the detector configuration, and
that a trigger based on the difference of arrival time of hits within
$\delta t=50\,$ns catches most of the events ($>$95\%). We use this
value for all detector configurations. 
For the $1+1$ mode, we optimized the cut on the time difference window
$\delta t=t_3-t_1$ ($t_{1,\, 3}$ are resp.\ the arrival times at the
lower  and upper sensor) in terms of $d^{5\sigma}$, and rescaled $n_s$
by the fraction of events surviving the time cuts $\delta
t=50,\,100,\,150\,$ ns, resp.\ for DDC's, DC and IC detector configurations, accordingly.

As we pointed out earlier, the distance reach with
DDC$_{4\pi}^{\rm{VL}}$ and sensor noise reduced to about
$100\,$Hz has a reach larger than $2\,$Mpc in the $2+1$ mode, using
the benchmark luminosity of \cite{KistlerAndBeacom}, twice the IC
Q.E., and modules with 50\% larger radius. It is eventually limited by
the signal statistics and not by the noise, so that it constitutes an
upper limit of this $>0.5\,$Mton ($2+1$ mode) configuration. In order
to perform routine SN detection, it is necessary to have $d^{5\sigma}
>4\,$Mpc. The detector extension must therefore be significantly more
ambitious and provide at least an effective volume of $2\,$Mton,
assuming that the noise is not limiting the reach. For an IceCube
extension that would mean the installation of $\gtrsim 10'000$ highly
efficient modules, probably with a very small string spacing, in order
to collect a substantial amount of events with hits spreading over
more than one string. Reducing further the spacing between sensors
slightly improves the reach, e.g.\ to $2.2\,$Mpc with $2\,$m, $3\,$m
being quite close to optimal w.r.t.\ the studied signatures. In
conclusion, the realistic implementation  of a supernova detector with a
reach up to $4\,$Mpc seems quite demanding from the point of view of
costs, deployment time, complexity and R\&D, but not unmotivated given
the great physics return. Not to mention the improvements it would
bring for neutrino studies in the $1 - 100\,$GeV range (oscillations,
indirect search for dark matter, southern sky and low energy neutrino
sources).

In conclusion, other projects such as a $0.5\,$Mton
Hyper-Kamiokande~\cite{Nakamura:HyperK}, or the aforementioned
$100\,$kton LENA \cite{LENA} would have a quite limited
potential as well w.r.t.\ the detection and characterization of
spectral neutrino light curves from extra-galactic supernovae. We
therefore confirm that a large multi-Mton effective volume is
necessary to this purpose, such as the 5 Mton DEEP
TITAND~\cite{Suzuki:TITAND} project
discussed in \cite{KistlerAndBeacom} and designed for routine
detection of supernova explosion at a rate of the order 1/yr.

\begin{table}[h]
\begin{tabular}{cc|ccc}
conf. \,&\, $kk'$ & $\tilde{V}_{\mathrm{eff}}^{\rm{Sf}}$ \,&\,
$\tilde{V}_{\mathrm{eff}}^{{\rm{FD,\,}}T=5\,\rm{MeV}}$ \,&\,
$\tilde{V}_{\mathrm{eff}}^{{\rm{FD,\,}}T=6.5\,\rm{MeV}}$
\\
\hline
\multirow{4}{*}{IC} & 1+0 & 764 & $1.03\,10^3$ & $1.32\,10^3$  \\
				& 2+0 & 7.4 & 11.5 & 16.4  \\
				& 1+1 & 1.67 & 3.00 & 4.83  \\ 
				& 2+1 & 0.12 & 0.25 & 0.45  \\
\hline
\multirow{4}{*}{DC} & 1+0 & 990 & $1.33\,10^3$ & $1.7\,10^3$  \\
				& 2+0 & 10.8 & 16.8 & 24.0  \\ 
				& 1+1 & 7.0 & 12.2 & 19.0  \\
				& 2+1 & 0.98 & 1.90 & 3.20  \\
\hline
\multirow{4}{*}{DDC} & 1+0 & 990 & $1.33\,10^3$ & $1.7\,10^3$  \\ 
				& 2+0 & 10.8 & 16.8 & 24.0 \\ 
				& 1+1 & 12.5 & 20.6 & 30.8  \\ 
				& 2+1 & 2.62 & 4.62 & 7.24 \\
\hline
\multirow{4}{*}{DDC$_{4\pi}$} & 1+0 & $2.8\, 10^3$ & $3.8\,10^3$ & $4.8\,10^3$  \\ 
				& 2+0 & 40.7 & 63.0 & 90.0  \\ 
				& 1+1 & 58.9 & 93.2 & 135  \\
				& 2+1 & 13.9 & 23.4 & 35.2  
\end{tabular}
\caption{Comparison of the sensor effective volume in $\rm{m}^3$ for
  Sf and Fermi-Dirac ($T=5$ and $T=6.5\,$MeV) spectra for the various
  detector configurations and detection modes.\label{table-Spectra}}
\end{table}

\subsection{Energy spectrum}
\begin{figure}[h]
  \centering
  \includegraphics*[width=0.45\textwidth,angle=0,clip,trim=0mm 0 0 0]{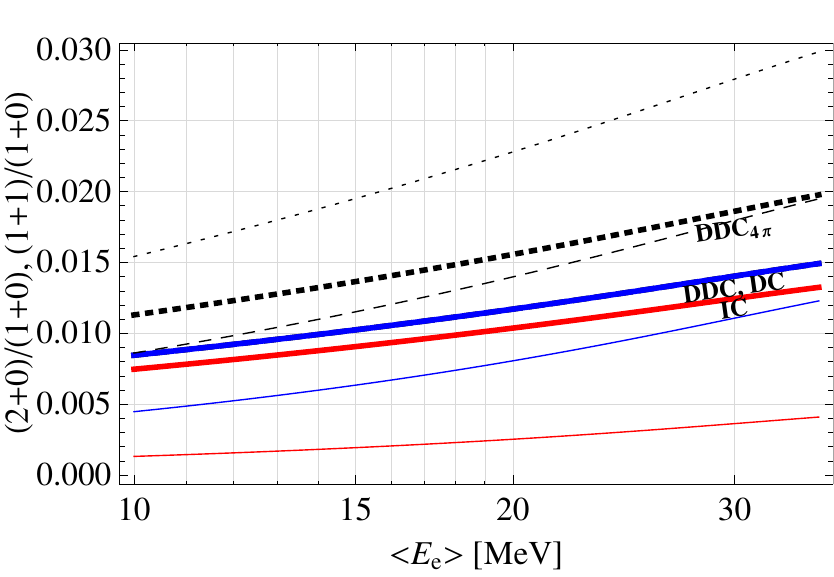}
  \caption{Coincidental hit ratios $(1+1)/(1+0)$ (thin lines) and
    $(2+0)/(1+0)$ (think lines) for DDC$_{4\pi}$, DDC, DC and IC, and
    w.r.t.\ $\langle E_{\mathrm{e}^{+}}\rangle$ (the same $\alpha$ was
    chosen for all $\langle E_{\mathrm{e}^{+}}\rangle$).}
  \label{fig:coincHitRatio}
\end{figure}

The hit ratios $(1+1)/(1+0)$ and $(2+0)/(1+0)$ are illustrated in
Fig.~\ref{fig:coincHitRatio}. Both observables are sensitive to the
average positron energy and do exhibit a 4.7\%/MeV variation around
the Garching energy for IC. All configurations exhibit equally strong
variations. In addition to their sensitivity to the spectral
parameters of the emission spectrum, observables based on ratios have
the advantage of canceling some of the involved systematics.

For the benchmark Garching model, Fig.~\ref{fig:Espectra} provides the
mean values for the neutrino spectrum, the positron spectrum and for
IC, for the various coincidental modes. It is noted that the mean
energy increases with the mode order.

The resolution on the spectrum will depend on the accumulated
statistics and can easily be calculated from the $n_s$ values
presented in Table~\ref{table-SF} and the distance. Considering
$\tilde{r}^{10}_{11}$ ($N_{10}\gg N_{11}$), the statistical
uncertainty is approximately $1/\sqrt{N_{11}}$. Assuming a spectrum
obeying Sf1 model with a slight uncertainty on $\langle E_\nu
\rangle$, we can estimate $\Delta \langle E_\nu \rangle$ from the
measurements, 
\[ \Delta \langle E_\nu \rangle  = \frac{1}{{\rm{d}}
  \tilde{r}^{10}_{11} /{\rm{d}}E (\langle E_\nu \rangle) \sqrt{N_{11}}
}.\]

For example, the statistical error on the fraction $1+1$/$1+0$ of the signal
recorded in the IC configuration for a benchmark SN assuming the Sf
spectral shape and located at the galactic center is about 5\% and
does therefore correspond to a resolution on the average energy of
about an MeV (and even $0.2\,$MeV for DDC$_{4\pi}$).

This result is extremely promising and deserves a more sophisticated
analysis beyond the scope of this paper in order to
study the potential for characterizing the spectral shape of the
neutrino flux, and possibly its full evolution, {\it i.e.} the
potential for obtaining the spectral neutrino light curve following a
galactic SN.

\subsection{Neutrino flux directionality} 
In the case of a galactic SN, it is desirable to locate the provenance of the SN 
neutrinos to inform the community, which could immediately
concentrate various instruments toward this target to observe the delayed
optical burst. Moreover, it is likely that some supernovae are
screened and never detected optically -- pointing to the target in
advance would result in a large multi-messenger sensitivity gain.

The underlying idea relies on the combination of

\begin{figure}[h]
  \centering
  \includegraphics*[width=0.5\textwidth,angle=0,clip,trim=0mm 0 0 0]{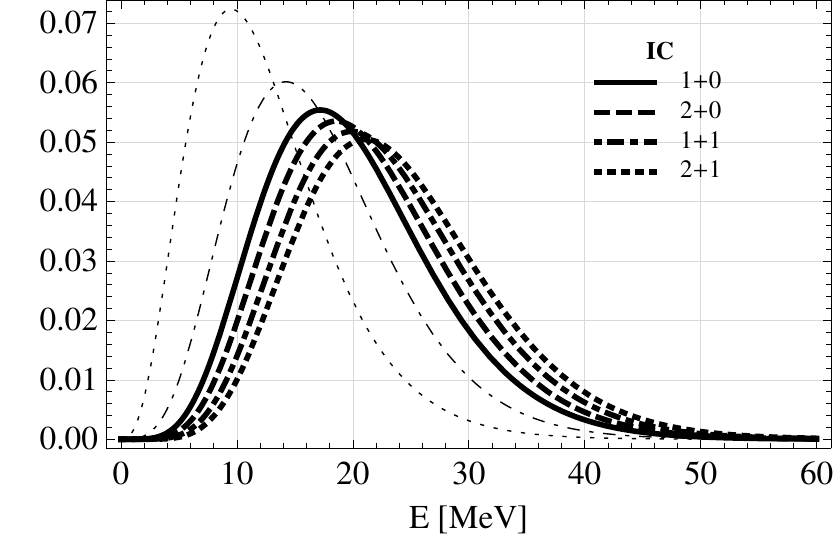}
  \caption{From left to right. Neutrino (thin, dotted) and positron
    spectra (Eq.~\ref {eq_positronSpectrum}): produced (thin,
    dot-dashed), producing single hit (solid), double hit (dashed),
    $1+1$ (dot-dashed) and $2+1$ (dotted) signatures for IC.
    The resp.\ mean energies (in MeV) are:
    \label{fig:Espectra}}
  \begin{tabular}{l|cccccc}
    &~~$\nu$~~&~~$\mathrm{e}^{+}$~~&~~$1+0$~~&~~$2+0$~~&~~$1+1$~~&~~$2+1$~~\\\hline
    ~~$\langle E \rangle$~~& 12.5 & 17.1 & 20.1 & 21.5 & 22.85 & 24.05\\
  \end{tabular}
\end{figure}

\noindent 1) the fact that the average positron energy leading to
coincidental hits is higher than for single hits, as illustrated in
Fig.~\ref{fig:Espectra}, which shows the detected positron spectra for
the various modes. There is a gradual increase of the average energy
of the modes with an increasing number of hits.

\noindent 2) an anisotropic scattering of the interaction with a
slight preference for photons emitted forward if the higher energy
tail of the neutrino emission can be selected. The average interaction
$\langle \cos{\theta} \rangle$ is quite small around SN average neutrino
energies. The neutrino interaction is forward scattered above
$\approx 14\,$MeV, which represents between 80 and 90\% of the detected
neutrinos depending on the coincidental detection mode).

$\langle \cos{\theta} \rangle = 1$\% in the $1+0$ detection mode and
rises to approx. 2\% in the $1+1$ detection mode with
IC~\cite{vissani-strumia} (see the average energies in
Fig.~\ref{fig:Espectra}). In order to exploit this 1\% difference, the
statistical uncertainty on $\tilde{r}^{10}_{11}$ must be kept smaller
and requires therefore large statistics only available below about 2
kpc with IC and 1 kpc with DC (and the azimuth of the incoming
direction is not resolved). The situation improves significantly with
DDC$_{4\pi}^{\rm{VL}}$ which will provide directional information for
a SN exploding anywhere in the MW, moreover in a design with small
inter-string spacing, the azimuthal direction will be constrained as
well (our 1D framework must be extended in order to include
inter-string coincidences).

It is also clear that the time difference distribution of detected
coincidental hits, an observable independent from
$\tilde{r}^{10}_{11}$, should also depend on the angle of provenance
of the neutrino flux $\vartheta$.

A careful assessment and quantification of the resolution by studying
the impact of the incoming direction on the induced distortion on the
hit time difference distribution and the various coincidental hit
ratios requires Monte Carlo studies. Such a study should also take
into account the elastic scattering channel. Although we pointed out
above that the low cross section of this channel will limit its
contribution to the overall detection potential, the strong angular
correlation and forward peaked cross section might result in a slight
enhancement for the measurement of the incidence direction.

\section{Conclusions}
In order to confirm the simple analytical calculations for single hit
effective detection volume and to precisely study the new coincident
hit detection methodology and the new observables to which it grants
access, precise simulation should be performed. In particular, a
precise assessment of the angular resolution in locating the SN from
the asymmetric Cherenkov emission can be reliably studied as a
function of distance and which is only qualitatively stated in this
paper.

With such a tool, a dedicated work not focused only on IBD but
implementing other neutrino interactions, e.g.\ with oxygen, would
permit the precise assessment of the potential of the novel analysis
method for oscillations and matter effect, hierarchy and
deleptonization neutrino emission phase studies.

A significant increase in potential would be reached with the
imagined DDC$_{4\pi}^{\rm{VL}}$ IceCube extension, with sensors
combining several improvements such as a larger and uniform collection
surface and an increased photo-detection efficiency (which is of specific
interest as the distance reach scales linearly at first order with the
Q.E.\ for $1+1$ and $2+0$ detection modes). SN detection up to $2\,$Mpc
would be at hand with the foreseen potential boost of this
detector. Such an extension, considerably cost effective compared
to similar projects, would be motivated by a broader experimental
interest, vastly enhancing the detection potential of low
energy neutrinos, $E_\nu \gtrsim 1\,$GeV, connected to physics of neutrino
oscillations, dark matter neutrino signatures, astrophysical neutrinos
in particular from the southern hemisphere, etc~\cite{DeepCore:physics}.

To reach about $2\,$Mton effective detection volume, thus granting access to
SN at distance scales larger of about $4\,$Mpc and guarantee routine SN
detection, the DDC$_{4\pi}^{\rm{VL}}$ configuration is likely too
modest and the number of modules should rather be about 10'000. Also R\&D
and extensive simulations will be necessary in order to determine the
optimal module arrangement, the sensor technology (e.g.\ the use of
solid state detectors instead of PMT,
see~\cite{solidStateAPDdetReview} for a review), and other hardware
related parameters.



It is worth mentioning an interesting outcome of this analysis:
the method described here can be readily applied to existing and future 
IceCube data in the regular muon data acquisition mode: events of 
$\ge20\,\mu$s duration are recorded at a rate of about $>2$ kHz, 
{\it i.e.} representing a lifetime fraction of about 5\%, without the necessity 
of a dedicated data stream. Numbers provided in this paper can readily 
be rescaled to obtain the reach in the
baseline IceCube configuration and its DeepCore provided a
corresponding rescaling of the distance reach (by a factor of about
4). Concerning a dedicated data stream, we do not discuss here the
possible hardware implications for the realistic implementations of
the methodology, but it may well already be implementable within the
current IceCube data acquisition hardware.

The methodology is natural for operating~\cite{antares} 
and projected~\cite{km3net} water detectors, given a comparatively 
large noise rate (based on~\cite{vladi1}, we found in~\cite{vladi2} 
a discussion based on a similar approach).


We finally note that accounting for spin-flavor conversion and MSW
effect for neutrinos propagating outward the SN core \cite{Volpe:2007qx}
, the $\bar\nu_{\rm{e}}$ flux is shifted to slightly higher
average energy, improving the prospects of the coincidental hit
method.


\section*{Acknowledgment}
We acknowledge fruitful discussions with our colleagues in IceCube, notably L. K\"opke.
This work is supported by the Swiss National Research Foundation under the grant PP002--114800.


\section*{References}
\small{
  
}

\end{document}